ARTICLE

# Physics-Guided Multi-Fidelity Surrogate Modeling for Natural Supercavity Geometry of an Axisymmetric Conical Cavitator

**Vinh Pham[1], Van Tu Nguyen[2,*] and Warn-Gyu Park[1,*]**

[1]School of Mechanical Engineering, Pusan National University, Busan, South Korea
[2] Faculty of Applied Science, Ho Chi Minh City University of Technology (HCMUT), VNU-HCM, Ho Chi Minh City, Vietnam
*Corresponding Author: Warn-Gyu Park (wgpark@pusan.ac.kr) and Van Tu Nguyen (vtnguyen@hcmut.edu.vn)

**ABSTRACT:** Natural supercavitation behind conical cavitators generates a vapor-filled envelope that, when sustained over the full body of a high-speed underwater vehicles, reduces hydrodynamic skin-friction drag and enables performance levels unattainable in fully wetted flow. Accurate prediction of maximum cavity length and maximum cavity radius and local cavity profile as functions of cavitation number σ and cone half-angle $a_c$ is essential for both the static design and the real-time shape management of super-cavitating torpedo cavitators. Classical analytical models provide computationally inexpensive predictions but exhibit systematic disagreements in the practically important regime of moderate cavitation number and large angle, where slender-body assumptions begin to break down. Existing data-driven machine learning approaches offer higher accuracy but are constrained by the prohibitive cost of multiphase computational fluid dynamics (CFD) sweeps over the full design space. This study proposes a physics-guided multi-fidelity surrogate framework based on Extended Hierarchical Kriging (PG-EHK) that addresses this challenge by fusing the multiple competing analytical cavity formulas as non-level low-fidelity (LF) trend functions corrected by a sparse set of high-fidelity (HF) CFD data. It simultaneously incorporates the analytical models regardless of their local fidelity ordering, using Bayesian-estimated scaling factors to construct the trend function as physical baseline. The resulting surrogate maps operational condition inputs to cavity geometry responses using fewer than 20 high-fidelity CFD evaluations. Cross-validation on five independent hold-out CFD cases demonstrates that the proposed PG-EHK surrogate achieves 61% reduction in prediction error compared to the best-performing individual analytical approach.

**KEYWORDS:** Natural supercavitation, physics-guided multi-fidelity surrogate, extended hierarchical kriging, conical cavitator, cavitator design method.

## 1 Introduction

Supercavitation is a hydrodynamic phenomenon in which an underwater body traveling at high speeds experiences a local pressure drop below the liquid's vapor pressure, resulting in the formation of a large vapor cavity. This phenomenon has attracted considerable attention in various high-speed underwater applications, including hydrofoils, projectiles, propellers, and high-speed torpedoes. As the body speed increases beyond a critical condition, natural cavitation may initiate locally and subsequently develop into an unsteady cavity that evolves with the surrounding flow [1]. The cavity dynamics become even more complex and highly unsteady under specific operating conditions, such as near-free-surface motion and free-surface entry and exit, where the interaction between the cavity and the free surface strongly affects the cavity evolution and hydrodynamic loads[2,3]. Supercavitation represents a particular cavitation regime in which the vapor cavity grows sufficiently large to envelop a substantial portion, or even the entire body, traveling through the liquid. When the cavity completely encloses the underwater vehicle, the wetted surface area in direct contact with the liquid is drastically reduced. Consequently, viscous friction drag is significantly decreased, potentially enabling substantial drag reduction and higher vehicle speeds for a given propulsion capability. In the fully developed supercavitating regime, the viscous contribution to the total resistance can become relatively small, with the hydrodynamic resistance being dominated primarily by the cavitator and other non-viscous components. Supercavity can be generated through natural or artificial approaches and has been extensively investigated using theoretical, experimental, and numerical methods [4–6]. Owing to the relatively stable shape of a fully developed supercavity under appropriate operating conditions, analytical and empirical approaches have been widely developed to predict cavity characteristics and hydrodynamic performance[7].

In engineering, the cavitator mounted at the forebody serves as the critical design element for high-speed underwater vehicles and supercavitating torpedoes. While flat disk cavitators exhibit fixed drag characteristics, conical cavitators provide distinct operational advantages because their hydrodynamic drag and cavity geometry vary dynamically with the cone angle [8]. A key design challenge involves accurately predicting primary supercavity dimensions, specifically the cavity length $L_c$ and maximum diameter $D_c$, as functions of the cavitation number $\sigma$ and cone half-angle $\alpha_c$. However, analyzing this relationship remains challenging because no exact analytical theoretical solution exists that connects drag, cone angle, and cavitation number to 3D axisymmetric bodies [8,9]. The VA-111 Shkval torpedo, developed by the Soviet Navy in the 1970s, demonstrated this principle operationally, sustaining speeds exceeding 100 m/s by enclosing the vehicle body within a supercavity generated by a blunt disk cavitator [10]. Subsequent research has explored a practical cavitator design method for straight-running supercavitating torpedoes, which identify the optimal cavitator parameters that minimize total hydrodynamic drag while ensuring motion stability during straight level flight [11].

Classical analytical models rely on simplified physical assumptions to estimate supercavity geometries behind submerged bodies. Foundational formulations proposed by several authors [8,9,12–16], utilize slender-body theory or momentum conservation principles to approximate cavity profiles. For example, Logvinovich's model applies the principle of independence of cavity section expansion to evaluate section growth along the symmetry axis. While these classical formulations perform well under low cavitation numbers and small cone angles, systematic discrepancies arise under off-design conditions. Comparative studies demonstrate that Logvinovich's formulation tends to yield conflicting cavity length predictions relative to other theoretical models. Furthermore, for conical cavitators with larger cone angles or at higher cavitation numbers, potential theory and linear approximations fail to capture non-linear viscous turbulent flow behavior and pressure drag variations. Computational fluid dynamics (CFD) simulations can resolve these complex viscous flow structures with high accuracy[6,17–21]. However, solely relying on high-fidelity CFD for multi-variable parametric design sweeps incurs extreme computational costs.

To accelerate engineering design without compromising accuracy, surrogate modeling and data-driven methods have gained widespread adoption across physical domains. A Bayesian-optimized random forest model trained on combined experimental and CFD demonstrated high accuracy in predicting ventilated supercavitation length and classifying regime types behind a disk cavitator [22]. Mo et al [23] combines Bayesian-optimized XGBoost with SHAP (Shapley Additive Explanations) to predict and explain cavity water depth and length for chute aerators Ensemble learning methods including random forest, gradient boosting, and XGBoost were applied to cavity length prediction for hydraulic aerators [24]. A proper-orthogonal-decomposition-based reduced-order model was applied to predict cavitation flow fields under varying conditions [25]. Most recently, Chen et al. (2025) [18] proposed both data-driven and knowledge-driven Gaussian process methods for predicting ventilated cavity length from freestream velocity and ventilation rate using sparse water-tunnel data, explicitly incorporating a monotonic physics constraint as a regularizer. These studies share a common limitation: they address either ventilated supercavitation, disk cavitators, fixed geometry (without half-angle as a free variable), or single-fidelity frameworks requiring large training datasets.

Multi-fidelity methods fuse abundant, low-cost information with sparse, expensive high-fidelity data to construct accurate predictive surrogate models [26–30]. Behrou et al [31] combines low-fidelity homogenized media simulations with high-fidelity dehomogenized microstructures and Physics-Informed Neural Networks (PINNs) to predict fluid velocity and pressure fields. In aerodynamic surrogate modeling, Pham et al.[32] introduced an adaptive data fusion framework designed to handle non-uniform data structures. Expanding upon this framework, Pham et al. developed the Extended Hierarchical Kriging (EHK) method to incorporate multiple low-fidelity datasets. Unlike conventional Co-Kriging[33] or standard hierarchical Kriging (HK) [34], which require strictly nested sample locations across fidelity levels, EHK accommodates non-level low-fidelity datasets from varied models. By integrating disparate low-fidelity trend functions with sparse high-fidelity corrections, EHK substantially improves global prediction accuracy while reducing computational expense. EHK has been validated on full-aircraft aerodynamic database construction, outperforming standard Kriging, HK, and other state-of-the-art multi-fidelity surrogates across a range of analytical benchmarks and the engineering case study [35]. This established EHK methodology provides the foundation for bridging theoretical cavity models and high-fidelity numerical computations in the present study

Despite the convergence of these research threads, no prior study has applied a multi-fidelity surrogate framework that simultaneously fuses multiple competing analytical supercavity models as non-level low-fidelity (LF) trend functions with sparse high-fidelity CFD data. Such a framework is required to predict natural super-cavity geometry behind a cone cavitator with a variable angle. To address this gap, the present work develops a physics-guided EHK (PG-EHK) framework by enhancing the standard EHK method. By simultaneously incorporating multiple classical analytical supercavitation formulas as baseline knowledge of physical laws, the proposed PG-EHK framework significantly improves prediction robustness. Because these classical models share the same underlying physical foundation yet employ different closure assumptions, they yield systematically distinct predictions with no globally consistent accuracy ordering across the design space. This inherent non-hierarchical relationship renders them well suited as non-level LF inputs, making the PG-EHK framework the natural choice for their principled fusion with high-fidelity CFD data.

The proposed PG-EHK framework enables rapid, accurate, and physically consistent prediction of cavity profile across the full practical design space of (σ, $\alpha_c$, x) using only 20 high-fidelity CFD evaluations. This capability directly enables efficient cavitator design and supports real-time cavity shape management strategies in which $\alpha_c$ is modulated to maintain a target cavity length over the body of a supercavitating torpedo.

The main contributions of this study are as follows:

1. The first framework of physics-guided multi-fidelity surrogate model for natural super-cavity geometry that treats multiple competing classical analytical models as simultaneous non-level LF trend functions within the EHK framework, requiring fewer than 30 HF evaluations
2. A comparative validation against each individual analytical model, demonstrating the surrogate's superiority across the design space $\sigma \in [0.05, 0.3]$, $\alpha_c \in [15°, 45°]$, $x \in [0,\ L_c]$.

The remainder of the paper is organized as follows. Section 2 defines the problem statement, design variables, outputs, and operating regime. Section 3 describes the methodology: the analytical cavity models, the EHK framework, the CFD simulation setup, and the proposed PG-EHK construction procedure. Section 4 presents the results, surrogate accuracy assessment. Section 5 concludes the paper and outlines future work.

## 2 Problem Statement

The objective of this study is to construct a computationally efficient surrogate model that accurately maps the cavitator design variables of (σ, $\alpha_c$, x) to the primary geometric characteristics of a natural supercavity generated by conical cavitator in high-speed underwater flow. Figure 1 illustrates the definitions of a conical cavitator and its associated axisymmetric cavity geometry. The cavitator geometry is parameterized by the cone angle $\beta_c$ and half-angle $\alpha_c = 0.5\beta_c$ , with a fixed slant length, $L_{scl}$; $R_n$ denotes the base radius of the cone cavitator. The three primary outputs of interest are: (i) the normalized cavity length $\bar{L}_c = L_c/R_n$ ; (ii) the normalized maximum cavity radius $\bar{R}_c = R_c/R_n$ , and (iii) the normalized cavity profile $\bar{r} = r(x)/R_n$, which describes the radial extent of the cavity as a function of axial position $x$ and is required for body coverage assessment in torpedo design. The design space is spanned by three independent variables: (i) the cavitation number σ; (ii) the cone half-angle $\alpha_c$; and the axial coordinate $x$ measured along the symmetric axis of the cavity.

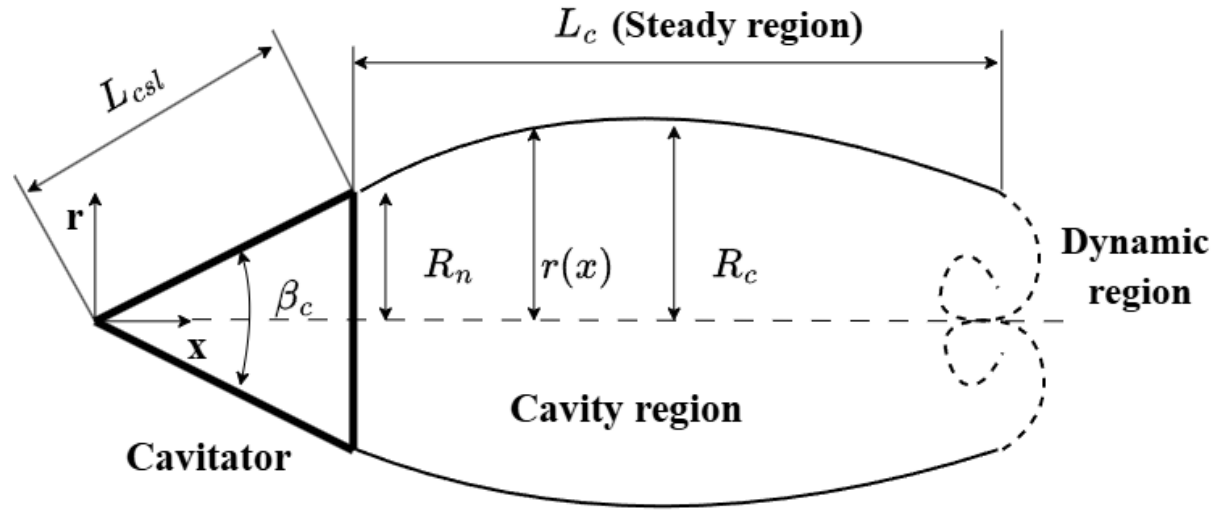

**Figure 1:** Definition of a conical cavitator geometry

The cavitation number is defined as:

$$\sigma = \frac{p_\infty - p_v}{0.5\rho V_\infty^2} \tag{1}$$

where $p_\infty$ is the freestream static pressure, $p_v$ is the vapor saturation pressure of water at the operating temperature, ρ is the liquid density, and $V_\infty$ is the freestream velocity.

The design space is bounded as $\sigma \in [0.05, 0.3]$, $\alpha_c \in [15°, 45°]$, and $x \in [0, L_c]$. The lower bound on σ corresponds to the deeply supercavitating regime of operational torpedo speeds; the upper bound corresponds to the onset of natural supercavitation. The bounds on $\alpha_c$ encompass the practical range of cone cavitator design, from slender cones that produce elongated narrow cavities to wide-angle cones with short, broad cavity envelopes.

The following operating regime assumptions apply throughout this study. First, the flow is treated as natural supercavitation. Second, the flow is axisymmetric, and the Froude number based on the cavitator diameter satisfies $Fr = V_\infty / \sqrt{(gD_n)} >> 1$, such that gravitational deformation of the cavity axis is negligible, and the cavity profile is also considered as axisymmetric. Third, the flow is incompressible (Ma << 0.1). Fourth, the Reynolds number (Re) based on $D_n$ satisfies $Re = V_\infty D_n/\nu \in [10^5, 10^6]$, which is the range representative of scaled torpedo experiments and full-scale torpedo operations. Under these conditions, Re effects on cavity geometry are secondary to σ and α effects.

The surrogate model takes the form:

$$\{\bar{L}_c, \bar{R}_c, \bar{r}\} = F(\sigma, \bar{\alpha}_c, \boldsymbol{\lambda}, \boldsymbol{\theta}, \xi) \tag{2}$$

Where $\bar{\alpha}_c = \alpha_c/180^0$ is the normalized cone half angle; $\xi = x/L_c$ is the normalized coordinate along the cavity length; $\boldsymbol{\lambda} = \{\lambda_1, \lambda_2, \dots, \lambda_{n_{LF}}\} \in \mathbb{R}^{n_{LF}}$ are the PG-EHK scaling factors for the $\boldsymbol{n_{LF}}$ analytical LF trend functions of cavity geometry; and $\boldsymbol{\theta} = \{\theta_1, \theta_2, \dots, \theta_m\} \in \mathbb{R}^m$ are the vector of model hyperparameters corresponding to *m*-dimensional input space. Both $\boldsymbol{\lambda}$ and $\boldsymbol{\theta}$ are estimated from HF training data from CFD experiment.

## 3 Methodology

### *3.1 Background*

#### *3.1.1 Analytical and Empirical Models for Supercavity Geometry*

Three distinct analytical approaches for modeling natural supercavity geometry, those of Guzebsky, Choi and empirical Logvinovich approaches, are incorporated as LF baseline models of prior physical knowledge in the proposed PG-EHK framework. All three are derived from the Logvinovich independence-of-section-expansion principle and the slender-body theory of axisymmetric cavity flows but differ in their closure assumptions and the treatment of the aft cavity region. Their shared physics foundation makes them approximately correlated but systematically distinct in predictions, satisfying the non-level criterion for EHK application.

First, the Guzebsky approach [14] provides approximation formulation for the main dimensions of stationary axisymmetric super-cavity of cone shaped cavitator and the cavitation drag coefficient of cones as follows:

$$\frac{D_c}{D_n} = \sqrt{\frac{c_d}{k\sigma}}\,, \quad k = \frac{1+50\sigma}{1+56.2\sigma}, \qquad \frac{L_c}{D_n} = \left[\frac{1.1}{\sigma} - \frac{4(1-2\bar{\alpha}_c)}{1+144\bar{\alpha}_c^2}\right]\sqrt{c_d \ln\frac{1}{\sigma}} \tag{3}$$

Where $D_n = 2R_n$ is the cone base diameter and $\bar{\alpha}_c = \frac{\alpha_c}{180^o}$. The estimate of drag coefficient is given in Guzebsky 's formula as:

$$C_d = C_{d0} + (0.524 + 0.672\bar{\alpha}_c)\sigma, \qquad 0 \le \sigma \le 0.25, \qquad 0 < \bar{\alpha}_c \le \frac{1}{2} \tag{4}$$

$$C_{d0} = 0.5 + 1.81(\bar{\alpha}_c - 0.25) - 2(\bar{\alpha}_c - 0.25)^2, \qquad \frac{1}{12} \le \bar{\alpha}_c \le \frac{1}{2} \tag{5}$$

$$C_{d0} = \bar{\alpha}_c(0.915 + 9.5\bar{\alpha}_c), \qquad 0 < \bar{\alpha}_c < \frac{1}{12} \tag{6}$$

Second, the Choi approach [8] proposes formulas to estimate the cavitation drag coefficient of cones as:

$$C_d = \frac{1.2\bar{\alpha}_C^{\pi/2}}{0.155 + \bar{\alpha}_c^{\pi/2}} - \frac{4\bar{\alpha}_c^{\frac{\pi}{4}}}{e^{5.5\bar{\alpha}_c}}\sigma + \sigma + \frac{C_{F0}}{tan\ \beta_c/2}\frac{(1-\bar{\alpha}_c)^3}{0.96 - \bar{\alpha}_c}(1+\sigma)^{0.4} \tag{7}$$

$$C_{F0} = \frac{0.073}{(logRe_{csl} - 2)^2} \tag{8}$$

$$Re_{csl} = \frac{V_\infty L_{csl}}{v} \tag{9}$$

Where $Re_{csl}$ is the Reynolds number of the cone based on $L_{csl}$, $V_\infty$ is the flow velocity and $v$ is the kinematic viscosity.

In Choi approach, the Garabedian formulas are also employed to estimate the maximum cavity length and radius as:

$$D_c = D_n\sqrt{\frac{c_d}{\sigma}}, \qquad L_c = \frac{D_n A}{\sigma}\sqrt{c_d}\,, \qquad A = \sqrt{ln\frac{1}{\sigma}} \tag{10}$$

Both Guzebsky and Choi approaches employ Logvinovich's analytical elliptical formula to estimate the cavity profile as:

$$R(x) = \sqrt{\frac{2x}{L_c}\left(2 - \frac{2x}{Lc}\right)(R_n^2 - R_c^2) + R_c^2} \tag{11}$$

Third, the Logvinovic‘s semi empirical approach [14] provides the direct estimation of maximum cavity length and radius and cavity profile as:

$$R_c = R_n\sqrt{3.659 + \frac{0.761}{\sigma}} \quad (12)$$

$$L_c = R_n(4 + \frac{3.595}{sigma}) \quad (13)$$

$$\frac{R}{R_n} = (1 + 3\bar{x})^{\frac{1}{3}}, \bar{x} \le 2 \quad (14)$$

$$\frac{R(x)}{R_n} = \sqrt{3.659 + 0.847(\bar{x} - 2) - 0.236\sigma(\bar{x} - 2)^2}, \bar{x} \ge 2 \quad (15)$$

Where $\bar{x} = x/R_n$

*3.1.2 Extended Hierarchical Kriging*

The purpose of the EHK method [36] is to integrate multiple non-level LF datasets simultaneously. Let $\boldsymbol{X} = \{x_1, x_2, \dots, x_m\} \in \Omega \subseteq \mathbb{R}^m$ denote the input vector and let $\{\hat{f}_{LF,i}(x),\ i \in \overline{1, n_{LF}}\ \}$ denote the $n_{LF}$ LF model predictions at $\boldsymbol{x}$, which are considered as baselines of physical knowledge. Let $\boldsymbol{y}_{HF}(X)$ denote the HF response.

The EHK framework combines an HF training dataset with multiple non-level LF models. The surrogate representation of the HF model takes the form:

$$\hat{f}_{HF}(x) = \boldsymbol{\lambda}^T \hat{\boldsymbol{f}}_{\boldsymbol{LF}}(x) + Z_d(x) \quad (16)$$

where $\hat{\boldsymbol{f}}_{\boldsymbol{LF}}(x) = \left[\hat{f}_{LF,1}(x), \dots, \hat{f}_{LF,n_{LF}}(x)\right]^{\mathrm{T}}$ is a vector of non-level LF models. Each LF is weighted by the unknown scaling factors $\boldsymbol{\lambda} = \left[\lambda_1, \lambda_2, \dots, \lambda_{n_{LF}}\right]^{\mathrm{T}}$, which collectively constitute the global trend component. $Z_d(x)$ is a stationary Gaussian random process with zero mean and covariance $\sigma_d^2$, representing the local discrepancy between the HF model and the ensemble of scaled LF models. Accordingly, the EHK prediction at any untried *x* can be obtained:

$$\hat{f}_{HF}(x) = \boldsymbol{\lambda}^T \hat{\boldsymbol{f}}_{\boldsymbol{LF}}(x) + \boldsymbol{c}(x)^T \boldsymbol{C}^{-1}\left(\boldsymbol{y}_{HF} - \boldsymbol{\lambda}^T \widehat{\boldsymbol{F}}_{LF}\right) \quad (17)$$

where $\widehat{\boldsymbol{F}}_{LF} = \left[\hat{\boldsymbol{f}}_{LF}\left(x_{HF}^{(1)}\right), \dots \hat{\boldsymbol{f}}_{LF}\left(x_{HF}^{(n_{HF})}\right)\right] \in \mathbb{R}^{n_{LF} \times n_{HF}}$ is the matrix of LF model responses evaluated at the HF training samples; $\boldsymbol{C}(x_{HF}, x'_{HF})$ is the correlation matrix among the HF training samples; $\boldsymbol{c}(x) = \phi(x, x_{HF})$ is a vector of the correlation between the untried point $x$ and each HF training sample; $\phi(x, x')$ is the spatial correlation function, dependent solely on the Euclidean distance between two points $x$ and $x'$; $\boldsymbol{\theta} = [\theta_1, \theta_2, \dots, \theta_m] \in \mathbb{R}^m$ is a vector of unknown hyperparameters governing the Gaussian correlation function.

$$\boldsymbol{C}(x_{HF}, x'_{HF}) = \left(\phi(x_{HF}, x'_{HF})\right)_{i,j} \in \mathbb{R}^{n_{HF} \times n_{HF}} \quad (18)$$

$$\phi(x, x') = \prod_{k=1}^{m} exp\left(-\theta_k \left|x^{(k)} - x'^{(k)}\right|^2\right) \quad (19)$$

To estimate the hyperparameters $\boldsymbol{\lambda}$, $\boldsymbol{\theta}$, and $\sigma_d^2$, the maximum likelihood estimation method is used to maximize the likelihood function given by

$$\boldsymbol{\theta} = \operatorname{argmax}[\ln\{L(\boldsymbol{\theta})\}] \tag{20}$$

where

$$ln\{L(\theta)\} = -n_{HF}\, ln(\sigma_d^2) - ln|\boldsymbol{C}| - \frac{\left(\boldsymbol{y}_{HF} - \boldsymbol{\rho}^T\widehat{\boldsymbol{F}}_{LF}\right)^T \boldsymbol{C}^{-1}\left(\boldsymbol{y}_{HF} - \boldsymbol{\rho}^T\widehat{\boldsymbol{F}}_{LF}\right)^T}{\sigma_d^2} \tag{21}$$

$$\lambda^*(\boldsymbol{\theta}) = \left(\widehat{\boldsymbol{F}}_{LF}{}^T \boldsymbol{C}^{-1}\widehat{\boldsymbol{F}}_{LF}\right)^{-1} \widehat{\boldsymbol{F}}_{LF}{}^T \boldsymbol{C}^{-1}\boldsymbol{y}_{HF} \tag{22}$$

$$\sigma_d^2(\boldsymbol{\theta}) = \frac{1}{n_{HF}}\left(\boldsymbol{y}_{HF} - \boldsymbol{\lambda}^T\widehat{\boldsymbol{F}}_{LF}\right)^T \boldsymbol{C}^{-1}\left(\boldsymbol{y}_{HF} - \boldsymbol{\lambda}^T\widehat{\boldsymbol{F}}_{LF}\right) \tag{23}$$

Note that the scaling factors $\boldsymbol{\lambda}$ are implicitly tuned during tuning of the hyperparameters.

### *3.2 Proposed Methodology*

The construction of the proposed PG-EHK model for the natural supercavity geometry behind a conical cavitator proceeds in four sequential steps, illustrated schematically in Figure 2. The detailed results of analysis are discussed in Section 4.

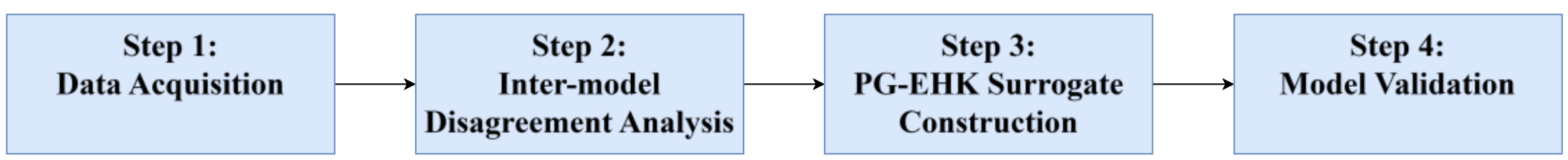

**Figure 2**: Flowchart for constructing the proposed PG-EHK surrogate model

- **Step 1** (Data acquisition**):** The performance of CFD and LF analytical analysis are performed on a uniform grid of N = 400 points over the design space ($\sigma \times \alpha_c \times \xi$: 5 × 4 × 20) to generate LF surfaces for $C_d$, $\bar{L}_c$, $\bar{R}_c$ and $\bar{r}(\xi)$. This evaluation of low-fidelity analytical analysis requires negligible computation time (< 1 s).

- **Step 2** (Inter-model disagreement analysis): The three traditional analytical approaches are compared to the CFD data to confirm that they exhibit non-level disagreement across the design space, justifying the PG-EHK approach over single analytical approach.

- **Step 3** (PG-EHK surrogate construction): The proposed PG-EHK surrogate is trained, incorporating the LF analytical models and CFD data.

- **Step 4** (Surrogate validation**)** Prediction accuracy is assessed using the K-fold cross-validation method and the following accuracy metrics of relative root mean square error (RRMSE) and relative maximum absolute error (RMAE) as:

$$RRMSE = = \frac{1}{STD}\sqrt{\frac{1}{n_{test}}\sum_{i=1}^{n_{test}}\left(y_i - \hat{f}_i\right)^2}\,, \qquad i \in \overline{1, n_{test}} \tag{24}$$

$$RMAE = \frac{1}{STD}\max\left|y_i - \hat{f}_i\right| \tag{25}$$

$$STD = \sqrt{\frac{1}{n_{test}} \sum_{1}^{n_{test}} (y_i - \bar{y}_i)^2} \tag{26}$$

Where and $\bar{y}_i$ and STD denote the mean and standard deviation of true responses over all test points, respectively; $n_{test}$ denotes the total number of testing points; $\hat{f}_i$ and $y_i$ are the predicted and true responses at the $\boldsymbol{i}$-th testing points, respectively;

*3.2.1 CFD Analysis: Generation of Training and Testing Data*

High-fidelity training and test data are generated using ANSYS Fluent. An axisymmetric computational domain featuring an axial length of $L_d = 71L_{csl}$ and a radial height of $H_d = 21L_{csl}$ is constructed, as illustrated in Figure 3. The conical cavitator is modeled as a solid boundary featuring a base slant length of $L_{csl} = 0.03$ m, a sharp apex, and variable half-angle $\alpha_c$. The domain is discretized using a structured mesh of 22891 cells, with boundary layer inflation applied at the cone surface; the first cell height is set to achieve a wall y+ ≈ 1 at the expected boundary layer thickness.

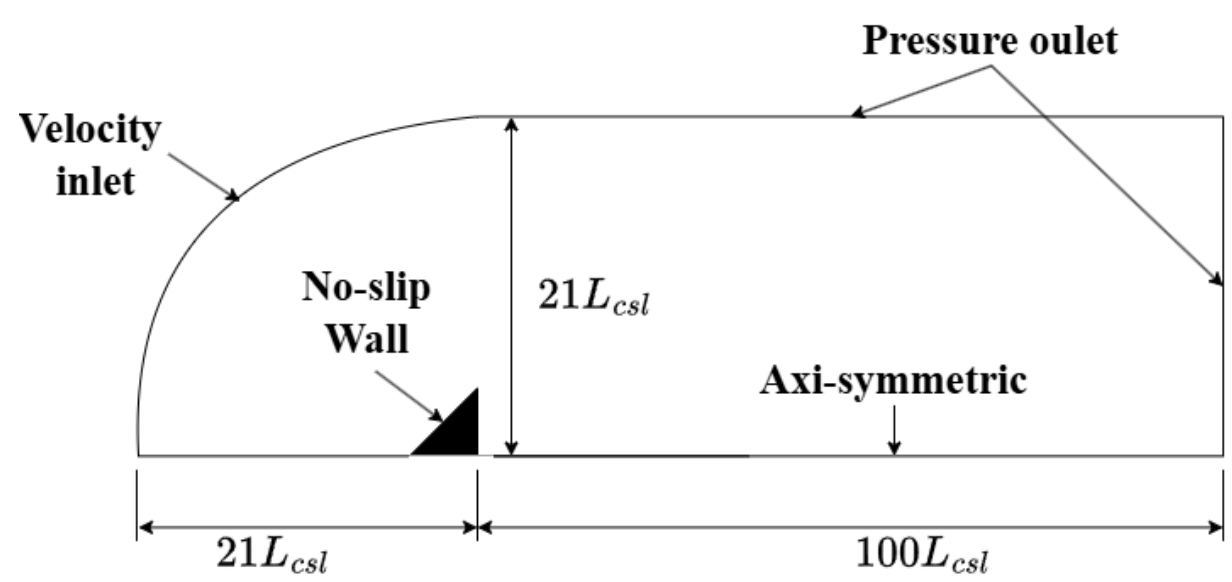

**Figure 3:** CFD computational domain and boundary setup

The following boundary conditions are applied: velocity-inlet on the left face; pressure-outlet at the right and top faces; no-slip wall on the cone surface; axis on the domain centerline; $p_\infty = 1.013 \times 10^5 Pa$, $p_v$ = 2340 Pa (saturation pressure at 20°C), ρ = 998.2 kg/m³ for all simulations.

The pressure-based solver is operated in transient mode with a fixed time step of $\Delta t = 1 \times 10^{-5}$ s, sufficient to resolve the dominant unsteadiness of the multiphase interface. The mixture multiphase model is employed in conjunction with the Schnerr–Sauer cavitation model to account for the transport of the liquid and vapor phases and the associated phase-change processes. The k-ω SST turbulence model is model is adopted because of its established capability to predict turbulent boundary layers and flow separation under adverse pressure gradients, which are important for resolving the boundary-layer development and cavity detachment in cavitating flows.

The sampling plan of CFD database spans three independent variables representing the simulated operational conditions of the cavitator: the cone half-angle, $\alpha_c$; cavitation number, $\sigma$; and the normalized axial coordinate, $\xi$ , measured along the centreline from the base of the cavitator to the distal end of the resolved cavity in each condition of ($\alpha_c$, $\sigma$). A full factorial design (FFD) is employed to generate a sampling plan encompassing 400 training points across the design space ($\alpha_c \times \sigma \times \xi$: 5 × 4 × 20), as shown in Table 1. The simulations were carried out on a computer featuring a configuration of an Intel® Xeon® W-2265 CPU@ 3.50 GHz, 12 cores, and 64 GB of RAM with 20 cases in total regarding conditions of $\alpha_c$ and $\sigma$. The computational cost is approximately 3 hours per case. It is noted that the initial flow conditions in the simulations are defined using the free-stream flow velocity, $V_\infty$ , as the direct input. The corresponding values of $\sigma$ can be conveniently estimated from $V_\infty$ using the Eq.(1).

**Table 1**: Sampling plan for CFD analysis and training database

| Variable | Range | Level | Unit |
|---|---|---|---|
| $\alpha_c$ | {15, 22.5, 30, 45} | 4 | deg |
| $\sigma$ | {0.3, 0.2, 0.15, 0.1, 0.05} | 5 | / |
| $V_\infty$ | {26, 32, 36, 44, 60} | 5 | m/s |
| $\xi$ | [0,1] | 20 | / |

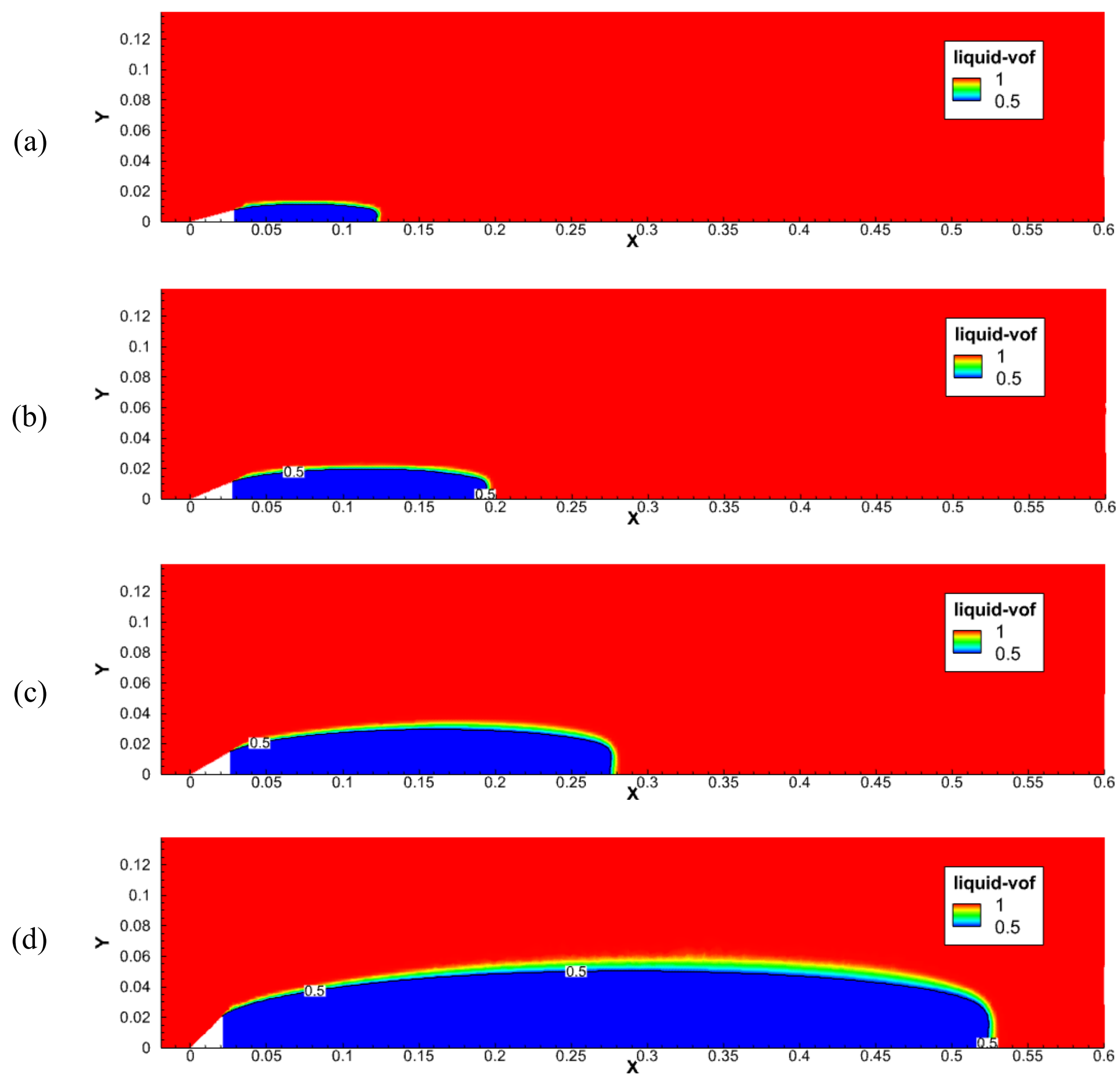

**Figure 4:** Cavity size from CFD analysis regarding flow velocity at $V_\infty = 44\ m/s$ ($\sigma = 0.1$) and configurations of conical cavitator: (a) $\alpha_c = 15^0$; (b) $\alpha_c = 22.5^0$; (c) $\alpha_c = 35^0$; (d) $\alpha_c = 35^0$.

The transient simulation proceeds until a statistically steady cavity shape is established. Convergence is declared when the cavity length $L_c(t)$, computed from the maximum axial extent of the $\alpha_v \geq 0.5$ iso-surface, stabilizes to within 1% variation over 200 consecutive time steps. The cavity boundary is defined as the iso-surface of vapor volume fraction $\alpha_v \geq 0.5$ iso-surface, stabilizes to within 1% variation over 200 consecutive time steps. The cavity boundary is defined as the iso-surface of vapor volume fraction $\alpha_v = 0.5$. The normalized cavity length $\bar{L}_c$ is extracted as the maximum axial extent of the $\alpha_v = 0.5$ region measured from the cone apex, divided by $R_n$. The normalized maximum cavity radius $\bar{R}_c$ is extracted as the maximum radial extent of the $\alpha_v = 0.5$ region, as shown in Figure 4. The final database constructed for PG-EHK training consists of 3 input variables of ($\alpha_c$, $\sigma, \xi$); and 4 responses of $C_d$, $\bar{L}_c$, $\bar{R}_c$ and sampled points of cavity boundary, $\bar{r}(\xi)$.

*3.2.2 Proposed PG -EHK Framework*

The PG-EHK framework for the surrogate modeling of axisymmetric natural supercavity geometries for an adaptive conical cavitator is illustrated in Figure 5. It is a hierarchical composite of EHK surrogates comprised of three sequential EHK layers that compute the drag coefficient, $\hat{C}_d$; normalized maximum cavity length and radius, $\hat{\bar{L}}_c$ and $\hat{\bar{R}}_c$, respectively; and normalized local cavity radius, $\bar{r}(\xi)$ as functions of the input variables $(\alpha_c, \sigma, \xi)$.

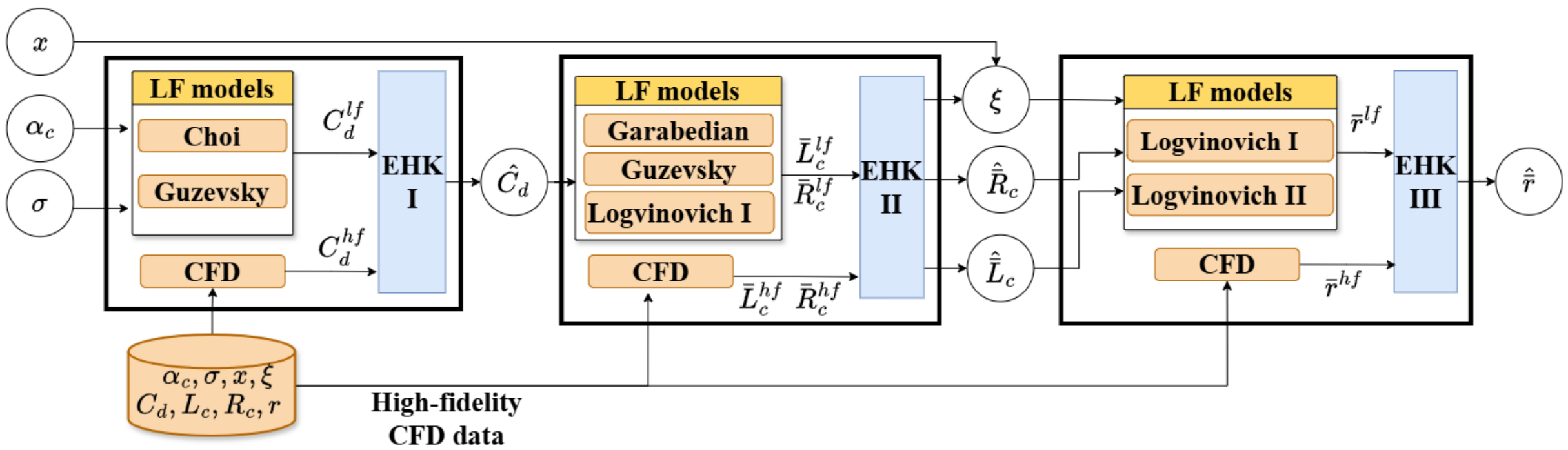

**Figure 5**. Architecture of PG-EHK model

In each EHK layer, the analytical formulas of Garabedian, Choi, Guzevsky, both empirical and analytical Logvinovich approaches are adopted as non-level LF models, together with high-fidelity CFD data, to construct a surrogate model for the quantity of interest. All EHK layers are trained independently, utilizing estimated data from the preceding layers:

- **Layer I** constructs surrogate of $\hat{C}_d$ with respect to the input of $(\alpha_c,\ \sigma)$. Three analytical drag coefficient formulas, those of the Guzevsky, Choi methods, given in Eqs. (4)-(8), are employed as LF models.
- **Layer II** constructs two surrogates, namely II-A and II-B, to estimate $\hat{\bar{L}}_c$ and $\hat{\bar{R}}_c$ as functions of inputs $(\alpha_c, \sigma, \hat{C}_d)$, respectively. Three analytical formulas of $\bar{L}_c$ and $\bar{R}_c$, namely those of Guzevsky, Garabedian and empirical Logvinovich (Logvinovich I) given in Eqs. (3), (10), (12) and (13), are employed as LF models.
- **Layer III** constructs surrogate of the local cavity radius, $\hat{\bar{r}}$, as function of input $(\alpha_c, \sigma, \hat{\bar{L}}_c, \hat{\bar{R}}_c, \xi)$. Two Logvinovich-based formulas, namely the analytical Logvinovich (Logvinovich II), given in Eqs. (11), and Logvinovich I, are employed as LF models.

## 4 Results and Discussion

This section presents the results and discussion of the proposed PG-EHK surrogate framework in two parts: (i) An inter-model disagreement analysis confirming the limitations of three complete analytical approaches, namely Guzevsky, Choi and empirical Logvinovich; and (ii) PG-EHK hyperparameter estimation and surrogate accuracy assessment. Both the Guzevsky and Choi approaches employ the analytical Logvinovich formula to estimate the cavity profile $\bar{r}(x)$, and differ only in the cavity length and radius that scale it. Therefore, the analytical Logvinovich approach of cavity profile appears hereafter in two distinct forms of Guzevsky and Choi approaches and rather than as a separate one.

### *4.1 Inter-Model Disagreement Analysis*

Figure 6–9 present a systematic comparison of the analytical approaches against CFD reference data for quantities of $C_d$, $\bar{L}_c$, $\bar{R}_c$ , and $\bar{r}(x)$, across four cone half-angles, $\alpha_c = \{15^o, 22.5^o,\ 30^o,\ 30^o, 45^o\}$; and a wide range of cavitation numbers, $\sigma = [0.05–0.30]$.

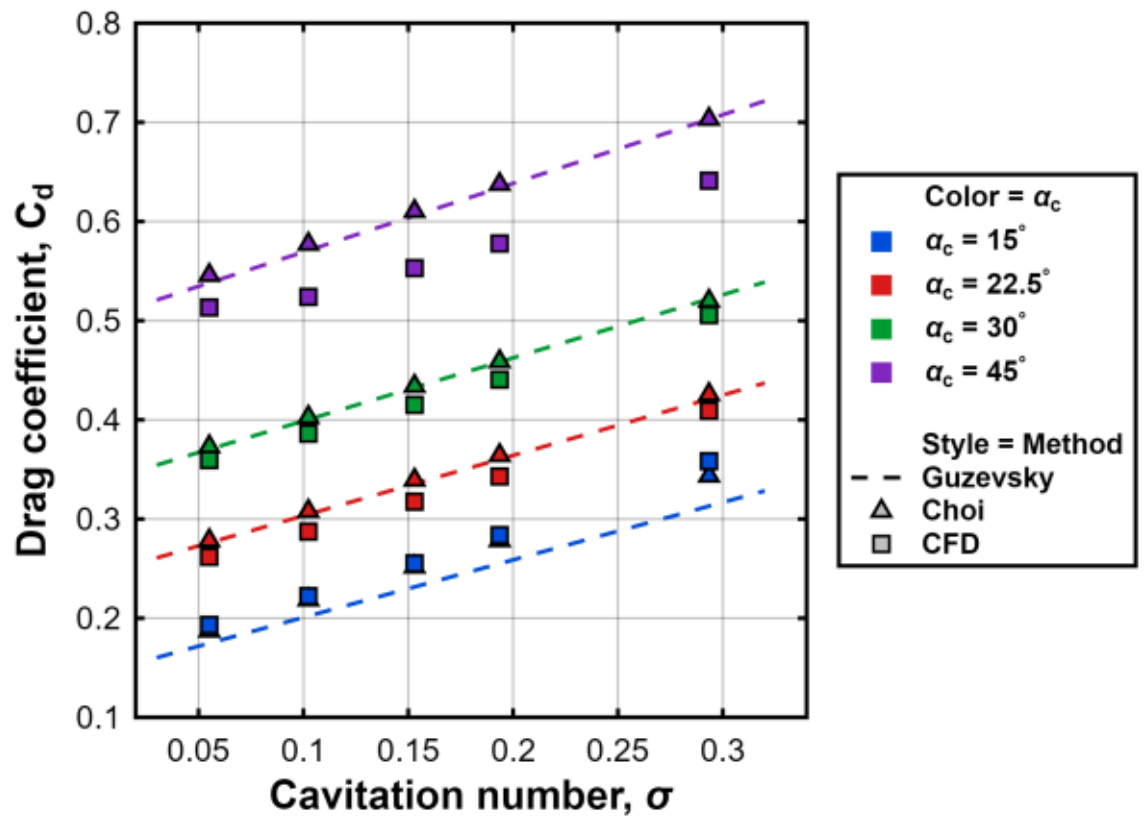

**Figure 6:** Comparison of the predicted $C_d$ regarding and $\sigma$ by analytical models vesus CFD reference across held-out cases of $\alpha_c = \{15^o, 22.5^o,\ 30^o,\ 30^o, 45^o\}$.

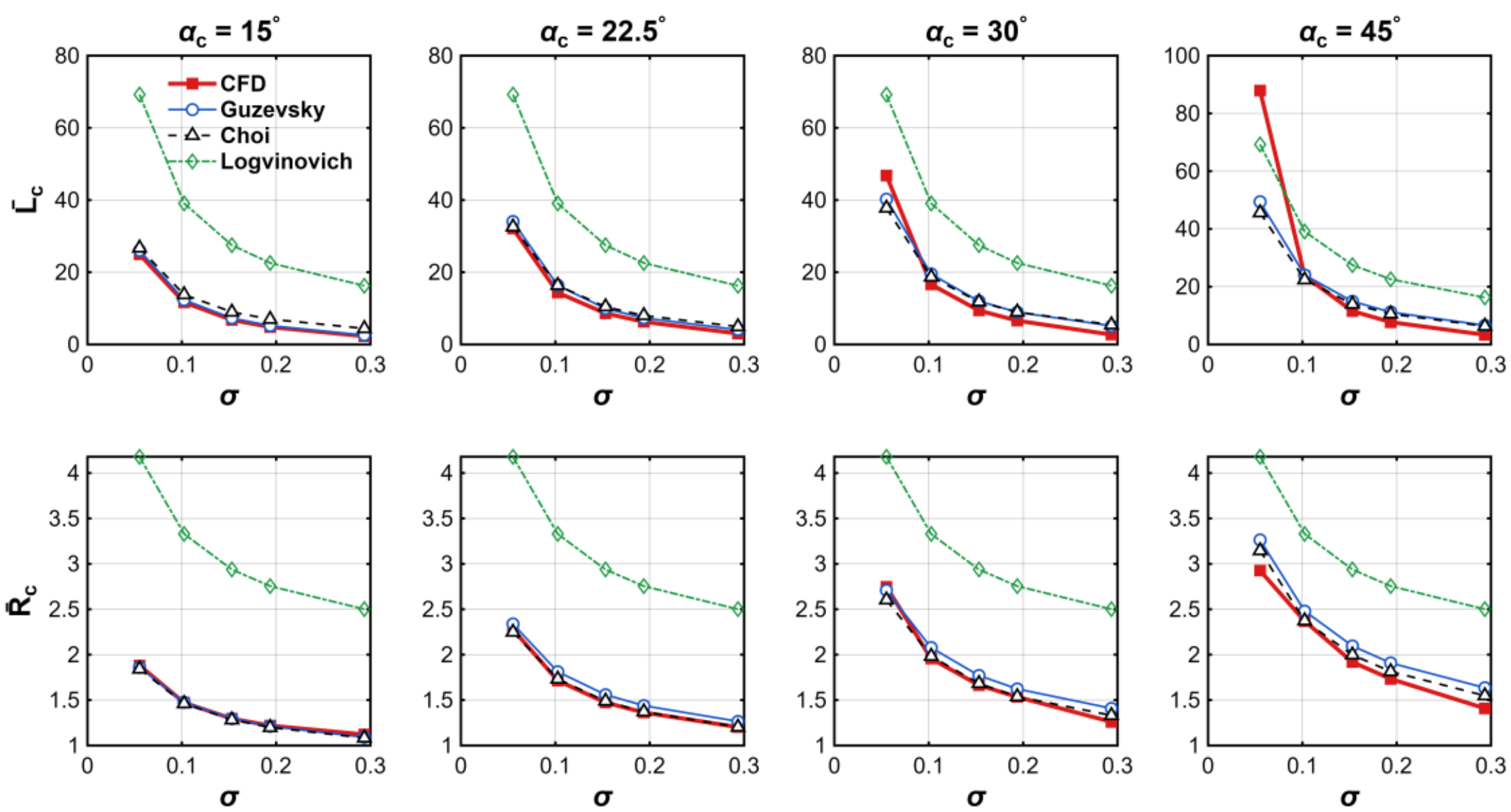

**Figure 7:** Predicted maximum cavity length $\bar{L}_c$ (upper row) and radius $\bar{R}_c$ (lower row) regarding $\sigma$ across held-out cases of $\alpha_c = \{15^o, 22.5^o,\ 30^o,\ 30^o, 45^o\}$.

Figure 6 present the drag coefficient, $C_d$, estimated from the analytical formulas in comparison with the CFD reference data. The four cone half-angles are distinguished by colors: $\alpha_c = 15°$ (blue); $\alpha_c = 22.5°$ (red), $\alpha_c = 30°$ (green); $\alpha_c = 45°$ (purple). Line and marker styles denote the data source: Guzevsky (dashed lines); Choi (filled triangles); CFD (filled square). Both analytical formulas and the CFD data consistently show that $C_d$ increases monotonically with $\sigma$ for all cone angles, and that larger $\alpha_c$ yields a higher drag coefficient at any given $\sigma$. This qualitative behavior is accurately captured in the Guzevsky and Choi formulas across the entire design space. Regarding quantitative accuracy, the two analytical formulas perform reasonably well at small cone angles ($\alpha_c = 15°$ and $22.5°$) where both formulas track the CFD data closely across the full range of $\sigma$. However, at larger cone angles ($\alpha_c = 30°$ and $45°$), a systematic overestimation emerges both Guzevsky and Choi formulas predict $C_d$ values consistently above the CFD reference, with the deviation becoming more pronounced at lower $\sigma$. The Choi formula tends to make stronger overestimate than that of Guzevsky at large $\alpha_c$, while at small angles the two formulas are nearly indistinguishable. Neither formula captures the nonlinear flattening of the $C_d$–$\sigma$ relationship observed in the CFD data at large cone angles and low cavitation numbers. The analytical formulas of Guzevsky and Choi formulas reproduce the qualitative trends of cavitator drag across the design space in a reliable manner, but exhibit a structured, configuration-dependent bias relative to the CFD data. The magnitude and sign of this bias vary systematically with both $\alpha_c$ and $\sigma$, rendering a simple constant correction factor insufficient to reconcile either model with the CFD reference. Since neither formula is uniformly superior, and since their prediction errors are smooth and spatially coherent, both are well suited to serve as physics-guided LF models within the first PG-EHK layer.

Figure 7 presents the normalized cavity length $\bar{L}_c$ (upper row) and normalized maximum cavity radius $\bar{R}_c$ (lower row) as functions of cavitation number $\sigma$, plotted for four cone half-angles, $\alpha_c = \{15°, 22.5°, 30°, 45°\}$, across four subplots. Four data sources are distinguished by styles: CFD (solid red line, filled squares); Guzevsky (solid blue line, open circles), Choi (dashed black line, open triangles), and empirical Logvinovich (dash-dot green line, open diamonds). All models and CFD data correctly reproduce the strong inverse dependence of both $\bar{L}_c$ and $\bar{R}_c$ on $\sigma$, with cavity dimensions decreasing sharply as $\sigma$ increases. The Guzevsky and Choi models track the CFD reference closely for $\bar{L}_c$ at small to moderate cone angles ($\alpha_c = 15°$–$30°$), with modest deviations remaining within acceptable bounds across most of the $\sigma$ range. At $\alpha_c = 45°$ and low $\sigma$, the CFD exhibits an abrupt surge in $\bar{L}_c$(exceeding 80) that neither Guzevsky nor Choi captures, indicating a physical nonlinearity that lies beyond the validity range of both slender-body models. The empirical Logvinovich model, by contrast, severely and consistently overestimates both $\bar{L}_c$ and $\bar{R}_c$ across all cone angles and cavitation numbers. The overestimation of $\bar{L}_c$ increases markedly at high $\sigma$, reaching ratios of 4–6 times the CFD value, while the $\bar{R}_c$ predictions are overestimated by factors of approximately 1.5–2.2 throughout the design space.

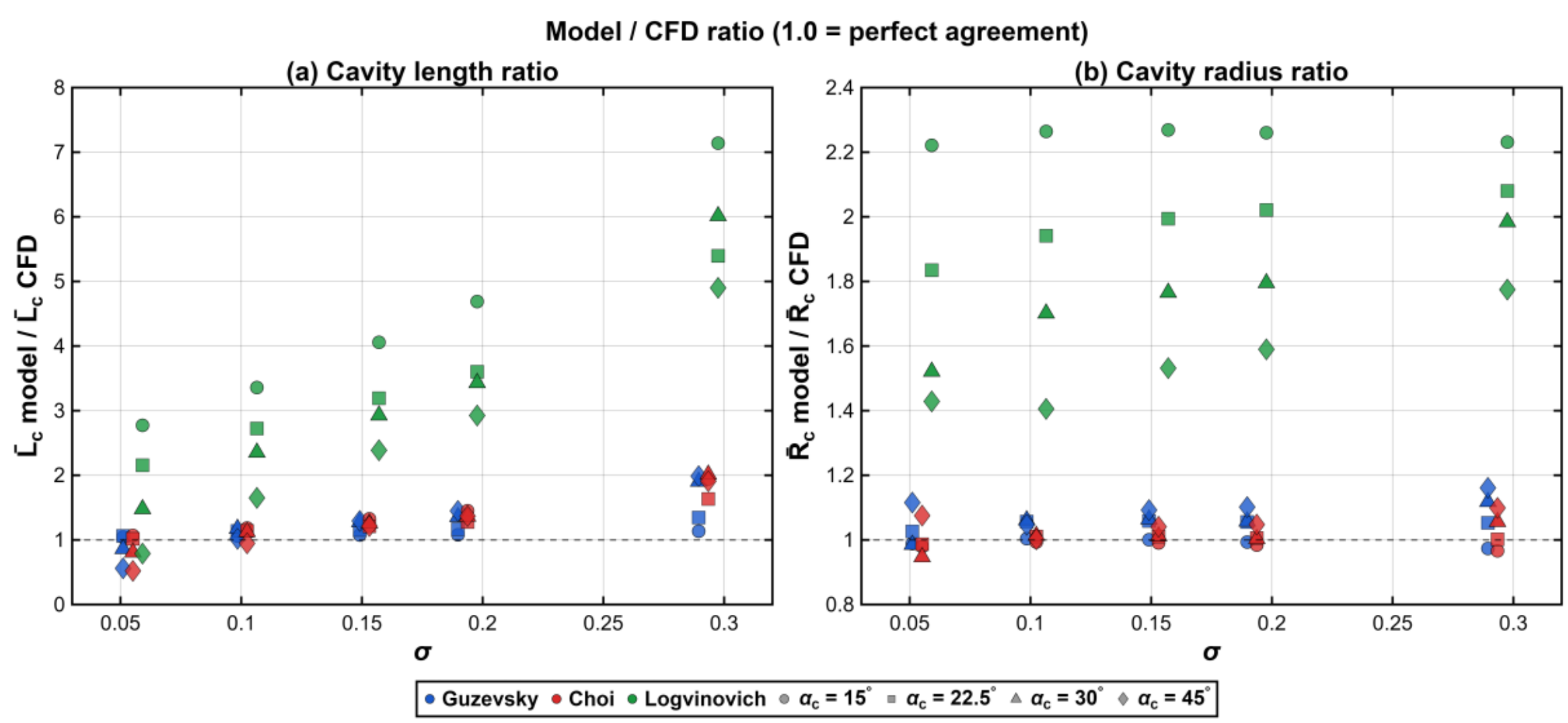

**Figure 8**: Model-to-CFD ratios regarding $\sigma$ for maximum cavity length ($\bar{L}_c^{model}/\bar{L}_c^{CFD}$, left) and maximum cavity radius ($\bar{R}_c^{model}/\bar{R}_c^{CFD}$, right) predicted by analytical models across held-out cases of $\alpha_c = \{15^o, 22.5^o, 30^o, 30^o, 45^o\}$.

In Figure 8, the ratio plots directly quantify the relative accuracy of each analytical model, with the dashed line at 1.0 indicating perfect agreement. Three data sources are distinguished by colors: Guzevsky (blue), Choi (red), and empirical Logvinovich (green). The four cone half-angles are distinguished by markers: $\alpha_c = 15°$ (circle); $\alpha_c = 22.5°$ (square), $\alpha_c = 30°$ (triangle); $\alpha_c = 45°$ (diamond). A ratio of 1.0 (dashed line) indicates perfect agreement with the CFD reference. For the cavity length ratio ($L_c^{model}/L_c^{CFD}$), the Guzevsky and Choi formulas cluster tightly around unity across all $\sigma$ values and cone angles, confirming their overall fidelity for $\bar{L}_c$ prediction. The Logvinovich formula diverges dramatically, with ratios climbing from approximately 1.4 at low $\sigma$ to values exceeding 6 at $\sigma = 0.3$, irrespective of cone angle. For the cavity radius ratio ($R_c^{model}/R_c^{CFD}$), a similar pattern is observed: Guzevsky and Choi formulas remain near unity, while that of Logvinovich overestimates $\bar{R}_c$ by factors of 1.3–2.2 across the entire domain.

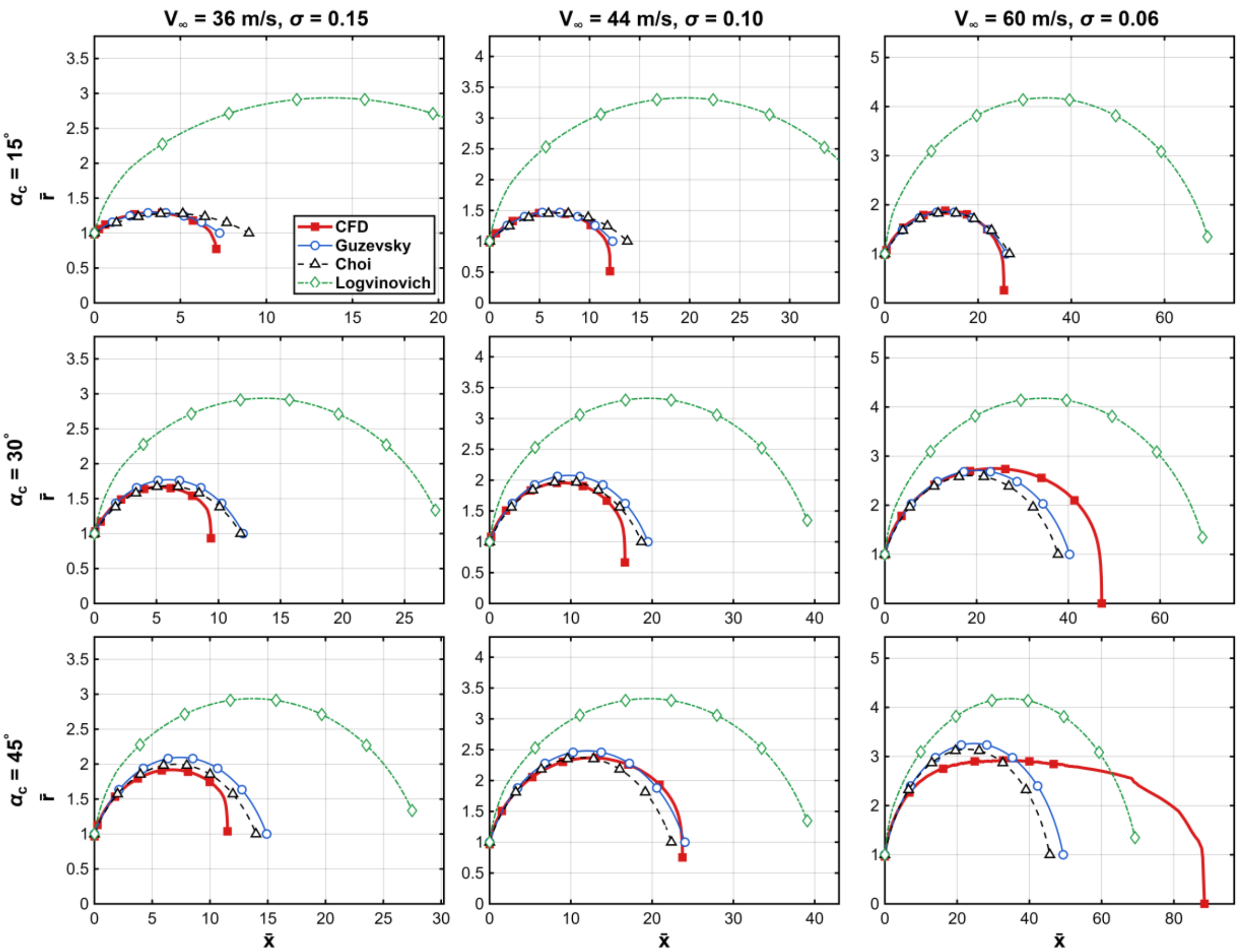

**Figure 9:** Predicted cavity profile, $\bar{r}(\bar{x})$, from analytical approaches and CFD reference at held-out cases regarding various $\alpha_c = \{15^o, 30^o, 45^o\}$, and $\sigma = \{0.15, 0.1,\ 0.06\}$.

The predicted local cavity profile comparisons in Figure 9 further illustrate the limitations of each analytical approach. Rows share the same $\alpha_c$, while columns share the same $\sigma$. Four data sources are distinguished by styles: CFD (solid red line, filled squares); Guzevsky (solid blue line, open circles), Choi (dashed black line, open triangles), and empirical Logvinovich (dash-dot green line, open diamonds). At $\alpha_c = 15°$, both Guzevsky and Choi approaches agree closely with the CFD profile shape for all three flow velocities, while empirical Logvinovich approach predicts a substantially larger and longer cavity that bears little resemblance to the CFD solution. As $\alpha_c$ increases to 30° and 45°, the profile predictions of both Guzevsky and Choi approach progressively diverge from the CFD reference, particularly in the rear closure region of the cavity, where the rate of radius decrease to axial distance is underpredicted by both formulas. At the highest velocity ($V_\infty = 60$ m/s, $\sigma \approx 0.05$) and largest cone angle ($\alpha_c = 45°$), the analytical approaches produce profile shapes that differ both in maximum radius and closure length, with Choi approach underestimating the rear cavity radius and Guzevsky approach failing to capture the correct closure slope. These discrepancies are consequential for engineering design, as inaccurate closure geometry directly affects vehicle drag, ventilation stability, and body-wetting predictions.

The collective evidence from the inter-model disagreement analysis reveals three critical limitations to the three existing analytical approaches:

- First, no single analytical approach is universally accurate across the full parameter space of $(\alpha_c, \sigma, \xi)$: both Guzevsky and Choi approaches are complementary in their error behavior, while the empirical Logvinovich approach carries a large, non-uniform bias for geometric quantities.
- Second, the inter-model spread between Guzevsky and Choi approaches themselves, which can exceed 20–30% for $\bar{L}_c$ at low $\sigma$ and large $\alpha_c$, constitutes a source of epistemic uncertainty that analytical methods alone cannot resolve without recourse to high-fidelity data.
- Third, the nonlinear dependence of cavity geometry on $(\alpha_c, \sigma)$ at extreme conditions is not captured by any of the closed-form expressions, all of which are derived from limiting assumptions (slender-body theory, potential flow, or empirical curve-fitting at specific design points). These findings collectively motivate the development of a physics-guided multi-fidelity surrogate framework that selectively combines the physics embedded in multiple analytical models as non-level low-fidelity inputs while using high-fidelity CFD data to correct their individual biases and extend predictive accuracy across the full design space.

## 4.2 PG-EHK Surrogate Construction and Accuracy Assessment

In this section, the proposed PG-EHK surrogate is compared against established analytical approaches, namely the Guzevsky, Choi, and empirical Logvinovich models, as well as a standard Kriging surrogate trained solely on CFD data, for the prediction of conical cavitator geometry across all combinations of $\alpha_c$ and $\sigma$. Both the PG-EHK and Kriging surrogates were trained on 15 CFD datasets, with 5 held-out cases reserved for model validation at $\alpha_c = 22.5^o$, and $\sigma$= {0.293, 0.194, 0.153, 0.102, 0.055}. The training phase of the PG-EHK and Kriging surrogates required 156 seconds and 118 seconds, respectively. Once trained, both models delivered inference in a few milliseconds on a single CPU core, which is negligible compared to the cost of a single CFD simulation.

**Table 2**: Estimated LF models' scaling factors in PG-EHK layers

| Layer | Output | Scaling factor | LF model | Value |
|---|---|---|---|---|
| I | $C_d(\alpha_c, \sigma)$ | $\lambda_{I,1}$ | Guzevsky | -1.41 |
| | | $\lambda_{I,1}$ | Choi | 2.29 |
| II-A | $L_c(\alpha_c, \sigma, C_d)$ | $\lambda_{IIA,1}$ | Guzevsky | 0.45 |
| | | $\lambda_{II-A,2}$ | Garabedian | 1.80 |
| | | $\lambda_{II-A,2}$ | Logvinovich I | -0.97 |
| II-B | $R_c(\alpha_c, \sigma, C_d)$ | $\lambda_{II-B,1}$ | Guzevsky | -1.17 |
| | | $\lambda_{II-B,2}$ | Garabedian | 2.21 |
| | | $\lambda_{II-B,3}$ | Logvinovich I | 0.018 |
| III | $R(\alpha_c, \sigma, L_c, R_c, \xi)$ | $\lambda_{III,1}$ | Logvinovich I | 0.040 |
| | | $\lambda_{III,2}$ | Logvinovich II | 0.962 |

The estimated scaling factors, $\lambda_i$, in Table 2 reveal how the PG-EHK framework weighs each LF formulas across the three layers, providing physical insight into the relative accuracy and bias of the underlying analytical formulas. In Layer I, the Choi formula receives the dominant positive weight ($\lambda_{I,2} = 2.29$) while the Guzevsky formula is assigned to a large negative weight ($\lambda_{I,1} = -1.41$), indicating that Guzevsky systematically overestimates $C_d$ and must be corrected downward by the surrogate. In models of II-A and II-B, the Garabedian formulas emerge as the most reliable LF source for cavity geometry, receiving the largest positive weights ($\lambda_{II-A,2} = 1.80$) for $\bar{L}_c$ and $\lambda_{II-B,2} = 2.21$ for $\bar{R}_c$). The Logvinovich I formula is assigned a strong negative weight for $\bar{L}_c$ ($\lambda_{3,2A} = -0.97$), consistent with its known tendency to substantially overestimate cavity length as observed in Figure 9. Its negligible contribution to $\bar{R}_c$ ($\lambda_{3,2B} = 0.018$) further confirms its limited utility for cavity radius estimation. In Layer III, the Logvinovich II formula dominates with $\lambda_{III,2} = 0.96$ , approaching unity, which demonstrates that the theoretical cavity profile shape is captured almost entirely by this formula alone, with the empirical variant adding only a minor correction ($\lambda_{III,1} = 0.040$). Collectively, these scaling factors confirm that the proposed PG-EHK framework selectively leverages complementary physics from each LF model while automatically suppressing the contributions of formulas known to carry systematic bias, without requiring any manual tuning of model weights.

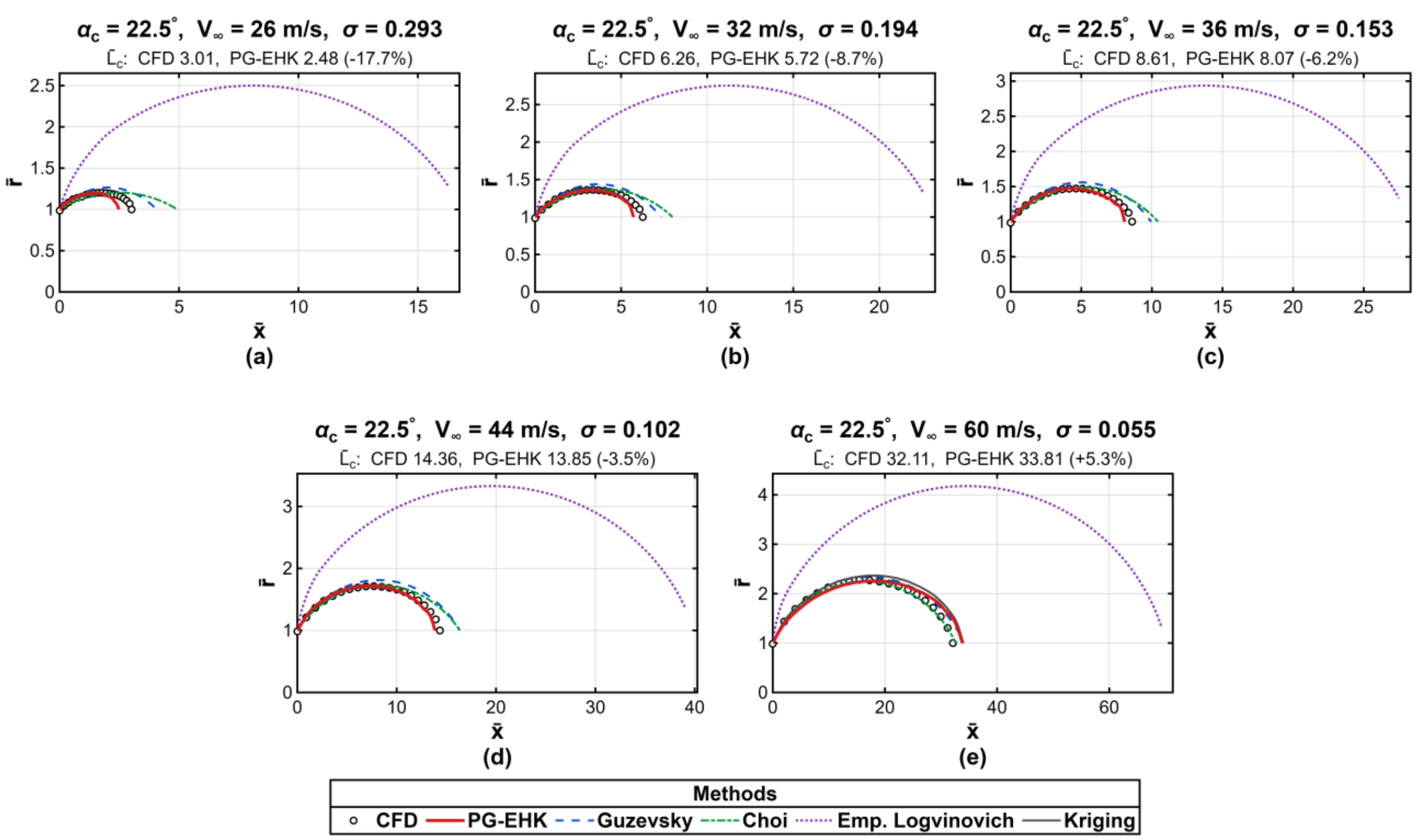

**Figure 10**: Predicted cavity profiles of models versus CFD reference on held-out cases of $\alpha_c = 22.5^o$: (a) $\sigma = 0.293$; (b) $\sigma = 0.194$; (c) $\sigma = 0.153$; (d) $\sigma = 0.102$; (e) $\sigma = 0.293$.

Figure 10 presents a comparison of the normalized cavity profiles predicted by the proposed PG-EHK and traditional approaches against CFD reference solution across five held-out test cases. All models are represented as follows: CFD (open circles); PG-EHK (red solid line); Guzevsky (blue dashed line); Choi (green dash-dotted line); Logvinovich I (purple dotted line); and Kriging (black solid line). The PG-EHK surrogates consistently produce the closest agreement with the CFD reference profiles across all five test cases, accurately capturing both the maximum cavity radius and the overall cavity profile shape. The prediction error in cavity length decreases progressively with decreasing $\sigma$, ranging from (−17.7%) at σ=0.293 to −3.5% at σ = 0.102, with a slight overestimation of +5.3% at the lowest cavitation number (σ =

0.055). By contrast, the Guzevsky and Choi approaches exhibit moderate but consistent deviations from the CFD data, particularly at low σ where cavity elongation is pronounced. The Kriging surrogates, while capturing the general cavity shape, underperform PG-EHK across most cases due to the absence of physics-based low-fidelity information. The empirical Logvinovich approach exhibits the largest prediction errors among all methods considered, consistently and substantially overestimating cavity length across the entire design space while yielding physically inconsistent cavity profiles. These deficiencies justify their exclusion from quantitative accuracy assessment.

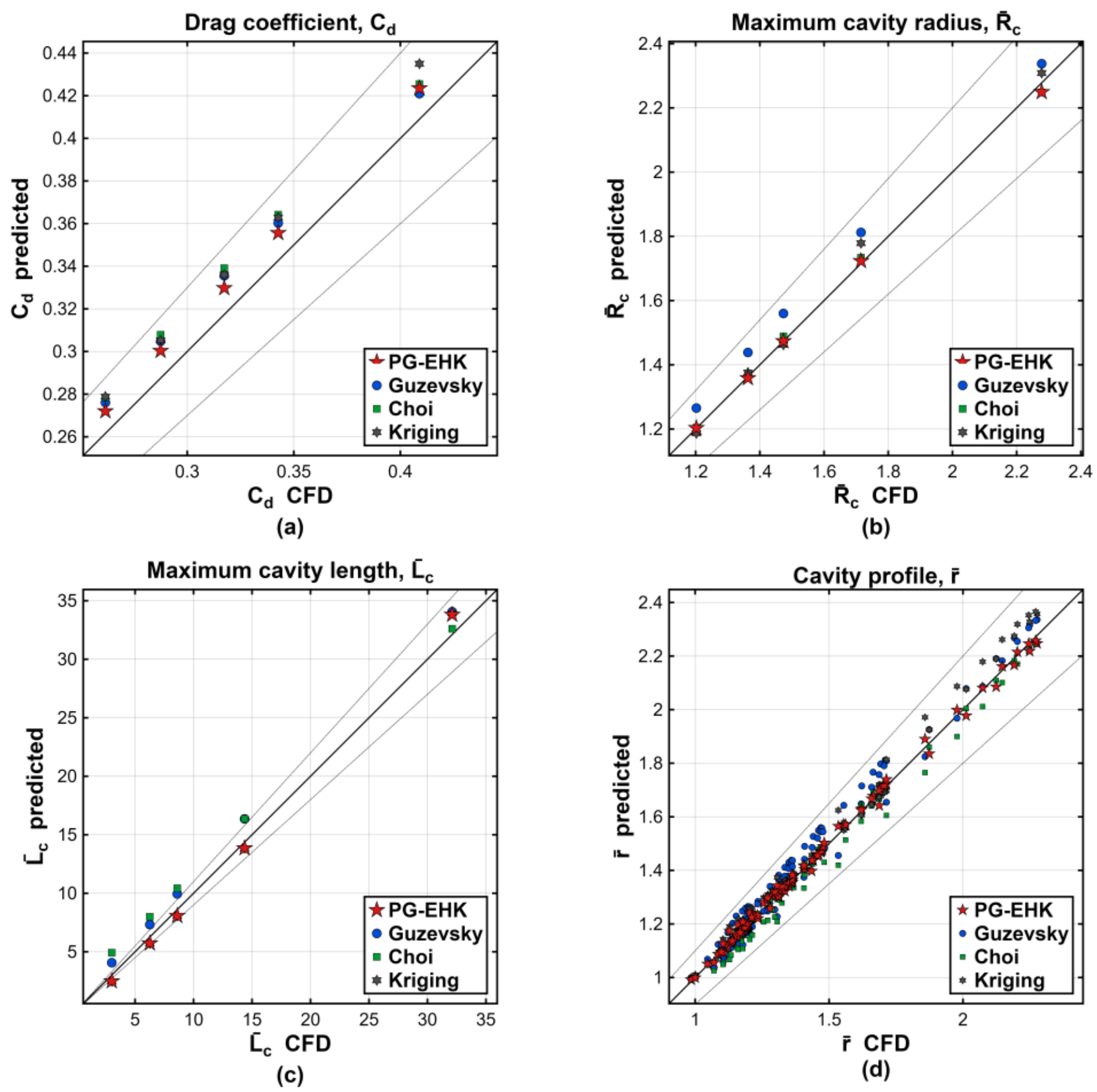

**Figure 11**: Scatter plots of PG-EHK predicted vs. CFD values on held-out set ($\alpha_c = 22.5^o$ , $\sigma$= {0.293, 0.194, 0.153, 0.102, 0.055}): (a) Drag coefficient, $C_d$; (b) Maximum cavity length $\bar{L}_c$; (c) Maximum cavity radius, $\bar{R}_c$; (d) Local cavity profile, $\bar{r}$.

Figure 11 presents parity plots of the predicted versus CFD-computed values for $C_d$, $\bar{R}_c$ , $\bar{L}_c$ , and $\bar{r}$ across the test dataset. All models are represented as follows: PG-EHK (red stars); Guzevsky (blue circles); Choi (green squares); and Kriging (black stars). PG-EHK's layers consistently exhibit the tightest clustering along the perfect-prediction diagonal in all four subplots, confirming its superior accuracy relative to the competing methods. For $C_d$, all models perform reasonably well, though Kriging produces a notable outlier at the highest drag value, indicating a loss of generalization beyond the training distribution. For $\bar{R}_c$ , Guzevsky approach systematically underpredicts across the mid-to-upper range, while PG-EHK model maintains close agreement with the CFD reference throughout. The $\bar{L}_c$ parity plot reveals that Guzevsky approach tends to overpredict cavity length at shorter cavities, whereas Choi and Kriging models exhibit moderate scatter; PG-EHK model remains the most consistent predictor across the full range. The cavity profile parity plot for $\bar{r}(x)$, which encompasses the largest number of data points, most clearly differentiates the methods: Guzevsky and Choi approaches display pronounced scatter and systematic underprediction at larger radius, Kriging shows moderate dispersion, while PG-EHK predictions remain tightly bounded within the error band along the diagonal. The empirical Logvinovich approach is excluded from this comparison, as its predictions deviate markedly from both the CFD reference data and the other analytical approaches, which would obscure the primary objective of validating the proposed PG-EHK model. These results collectively confirm that the proposed PG-EHK yields the most reliable and consistent predictions of supercavity geometry across all output quantities considered.

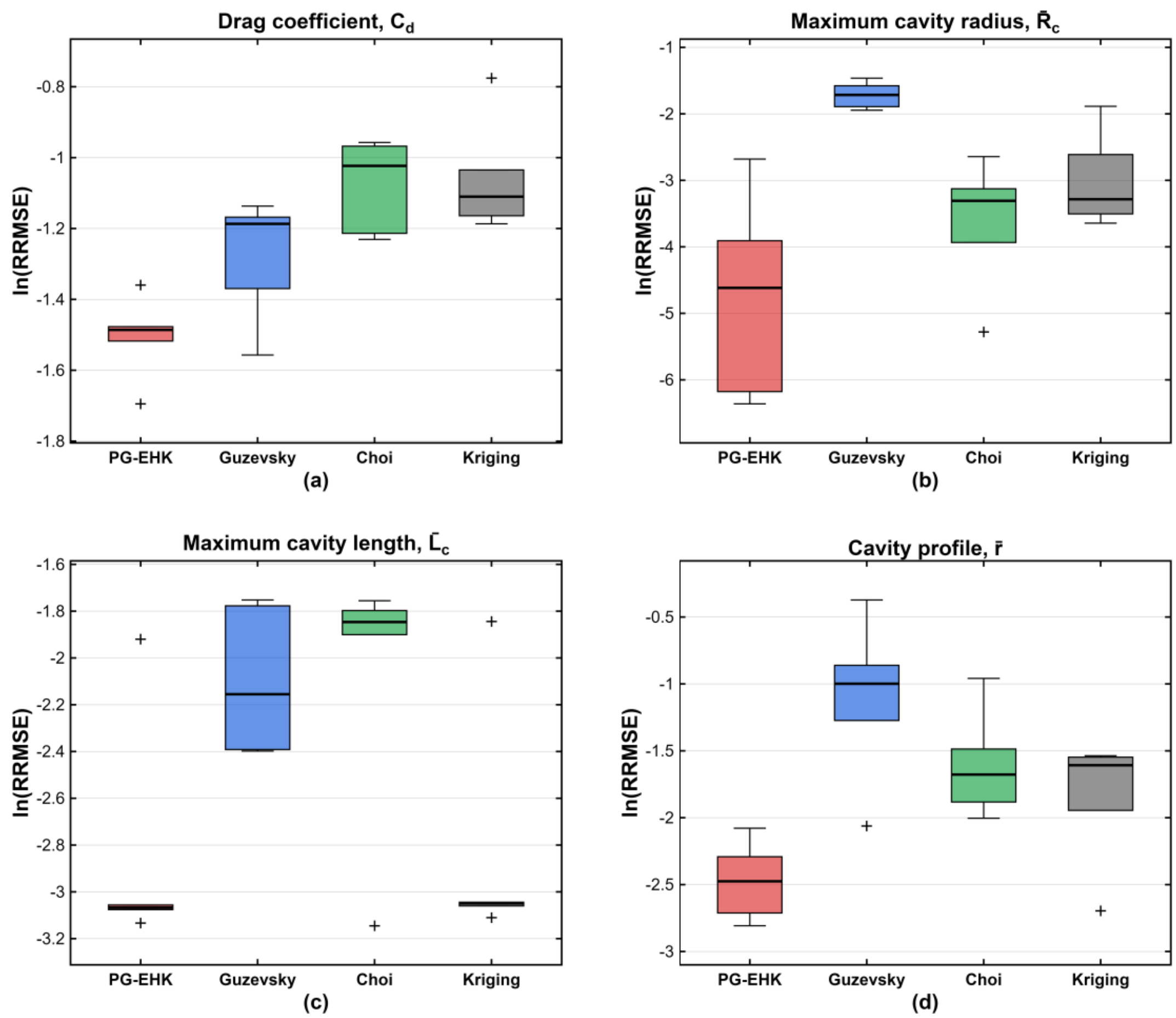

**Figure 12**: Accuracy comparison in RRMSE of prediction models against CFD on cross-validation test: (a) Drag coefficient, $C_d$; (b) Maximum cavity length $\bar{L}_c$; (c) Maximum cavity radius, $\bar{R}_c$; (d) Local cavity profile, $\bar{r}$.

For additional comprehensive model validation, a K-fold cross-validation method is employed to evaluate the performance of PG-EHK and Kriging models, in which 15 and 5 cases are randomly selected as the training and testing sets, respectively, from the original 20-case CFD data set. Figure 12 and Figure **13** compare the cross-validation results of the proposed PG-EHK surrogate against the analytical and Kriging models regarding the accuracy metrics of RRMSE and RMAE, respectively. The PG-EHK surrogate achieves the lowest RRMSE and RMAE for all output quantities. Specifically, PG-EHK is 21% more accurate than the best-performing analytical approach (Guzevsky) and approximately 37% more accurate than Kriging for $C_d$; 23% more accurate than the best-performing analytical approach (Choi) and approximately 59% more accurate than Kriging for $\bar{R}_c$ ; 42% more accurate than the best-performing analytical approach (Guzevsky) and approximately 6% more accurate than Kriging for $\bar{L}_c$; and 61% more accurate than the best-performing analytical approach (Choi) and approximately 49% more accurate than Kriging for $\bar{r}(x)$. Again, the results demonstrate that the proposed PG-EHK enables superior cavity profile prediction relative to both purely data-driven and purely analytical approaches.

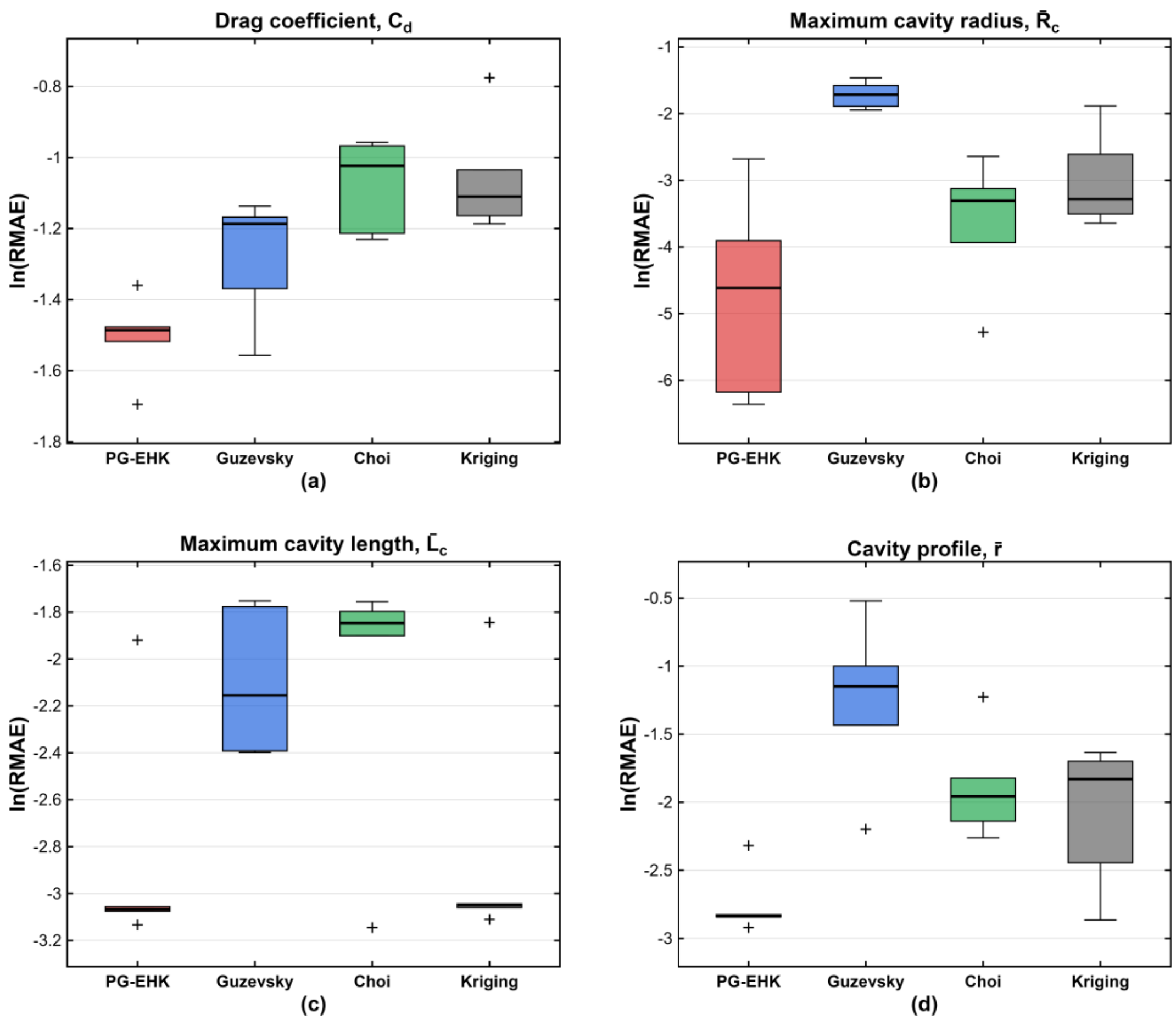

**Figure 13**: Accuracy comparison in RMAE of prediction models against CFD on cross-validation test: (a) Drag coefficient, $C_d$; (b) Maximum cavity length $\bar{L}_c$; (c) Maximum cavity radius, $\bar{R}_c$; (d) Local cavity profile, $\bar{r}$.

## 5 Conclusion

This paper has presented a novel physics-guided multi-fidelity framework for the surrogate modeling of axisymmetric natural supercavity geometries generated by an adaptive conical cavitator. The framework integrates multiple established analytical formulas as non-level low-fidelity models within a hierarchical three-layer EHK architecture, trained on a compact set of 20 high-fidelity CFD evaluations distributed over the design space of cone half-angle $\alpha_c$, cavitation number $\sigma$, and axial location $x$. The principal contributions and findings of this work are summarized as follows:

- The inter-model disagreement analysis demonstrated that no single existing analytical model is universally accurate across the full parameter space. The empirical Logvinovich model overestimates cavity length and radius by factors of 2–6 and 1.5–2.2, respectively, while the Guzevsky and Choi models, though more reliable for moderate cavitation numbers, exhibit non-systematic errors at low $\sigma$ and large cone angles that cannot be corrected by a simple analytical adjustment. These limitations collectively establish the need for a data-informed surrogate that can reconcile complementary analytical physics with high-fidelity CFD evidence.

- The proposed PG-EHK framework addresses this need by automatically learning optimal scaling weights for each analytical LF model from the CFD training data, as reflected in the estimated scaling factors reported in Table 2. The resulting surrogate achieves superior predictive accuracy relative to both the best-performing analytical model and a standard Kriging surrogate trained directly on CFD data alone, outperforming the best analytical model by 21%, 23%, 42%, and 61% for $C_d$, $\bar{R}_c$, $\bar{L}_c$, and $\bar{r}(x)$, respectively, and outperforming Kriging by 37%, 59%, 6%, and 49% for the same quantities. Parity plots and cavity profile comparisons on five held-out test cases confirm that PG-EHK reproduces the CFD cavity geometry with consistent fidelity across the full range of $\alpha_c$ and $\sigma$ considered, including the challenging low $\sigma$ regime where analytical methods fail most significantly.
- These results demonstrate that the physics-guided multi-fidelity architecture of PG-EHK enables accurate and computationally efficient supercavity geometry prediction from a small CFD training dataset, making it well-suited for integration into early-stage design workflows for supercavitating underwater vehicles.

Several directions for future work are identified. First, the current framework employs a fixed full-factorial training plan of 20 CFD evaluations; future work will investigate adaptive infill sampling strategies driven by the PG-EHK prediction variance, with the objective of achieving equivalent or superior surrogate accuracy with fewer than 15 CFD training points. By concentrating new CFD evaluations in regions of high predictive uncertainty, particularly at low $\sigma$ and large $\alpha_c$ where analytical models are least reliable, such an approach has the potential to reduce the computational cost of surrogate construction substantially while preserving accuracy across the full design space. Second, an inverse PG-EHK formulation will be developed to enable real-time, closed-loop cavity length control, in which the cone half-angle $\alpha_c$αc is modulated continuously to maintain a prescribed target cavity length $L_c$ against disturbances in the ambient cavitation number $\sigma$. The low evaluation cost of the trained surrogate makes it directly compatible with model-predictive control architectures operating at actuation timescales, offering a principled pathway from offline surrogate training to online feedback control of supercavity geometry. Third, the validated PG-EHK surrogate will be incorporated into a multidisciplinary preliminary design framework for the adaptive cavitator and body configuration of supercavitating torpedoes, addressing the coupled optimization of hydrodynamic performance, structural constraints, and cavity stability under realistic operational envelopes. These extensions will be reported in forthcoming publications.

**Acknowledgement:** This research was supported by Basic Science Research Program through the National Research Foundation of Korea (NRF), funded by the Ministry of Education (RS-2023-00248070), and was also supported under the framework of international cooperation program managed by National Research Foundation of Korea (RS-2026-16432970).

**Funding Statement:** This research was supported by Basic Science Research Program through the National Research Foundation of Korea (NRF), funded by the Ministry of Education (RS-2023-00248070), and was also supported under the framework of international cooperation program managed by National Research Foundation of Korea (RS-2026-16432970).

**Author Contributions:** The authors confirm contribution to the paper as follows: Conceptualization, methodology, software, formal analysis, data curation, validation, visualization, writing—original draft preparation, **Vinh Pham**; investigation, resources, writing—review and editing, **Van Tu Nguyen**; investigation, writing—review and editing, supervision, project administration, **Warn-Gyu Park**. All authors reviewed and approved the final version of the manuscript.

**Availability of Data and Materials:** Data available on request from the authors.

**Ethics Approval:** Not applicable.

**Conflicts of Interest:** The authors declare no conflicts of interest.

**Supplementary Materials:** Not applicable

**Abbreviations**

The following abbreviations are used in this manuscript:

| Symbol | Definition | Unit |
|---|---|---|
| $\widehat{(.)}$ | Quantity predicted by a surrogate or analytical model, | - |
| $\overline{(.)}$ | Quantity normalized by the cavitator base radius $R_n$ | - |
| $\alpha_c$ | cone half-angle of the cavitator | deg |
| $\bar{\alpha}_c = \alpha_c/180^o$ | normalized cone half-angle, as used in the drag correlations | - |
| $\beta_c = 2\alpha_c$ | Full cone angle | deg |
| $\sigma$ | Cavitation number | - |
| $\sigma_d^2$ | Process variance of discrepancy model | - |
| $\phi(.)$ | Correlation function, basic functions | - |
| $\mu$ | Correction factor | |
| $\lambda$ | Scaling factor | - |
| $\rho$ | Water density | kg/m$^3$ |
| $\theta$ | Hyperparameter | - |
| ν | Kinematic viscosity | m$^2$/s |
| $C_d$ | Drag coefficient | - |
| $\mathbf{c}(x)$ | Correlation vector | - |
| $\boldsymbol{C}$ | Correlation matrix | - |
| $D_n$, $R_n$ | Diameter and radius of cavitator | m |
| Fr | Froude number | - |
| $f(x)$ | Actual function | - |
| $\hat{f}(x)$ | Approximation function | - |
| $\hat{F}_{LF}$ | Matrix of responses of LF models at HF training points | - |
| g | Gravity | m/s$^{-2}$ |
| $L_c, \bar{L}_c = L_c/R_n$ | Maximum cavity length | m, - |
| $L_{scl}$ | Slant length of the conical cavitator | m |
| $n_{LF}$ | Number of LF models | - |
| $n_{test}$ | Number of testing points | - |
| $m$ | number of input dimensions | |
| $p_\infty$ | Ambient pressure | Pa |
| $p_v$ | Vapor pressure | Pa |
| $R_c, \bar{R}_c = R_c/R_n$ | Maximum cavity radius | m, - |
| $r(x), \bar{r}(x) = r/L_c$ | Local cavity profile | m, - |
| Re | Reynold number | - |
| $V_\infty$ | Velocity of free stream | m/s |
| $x, \bar{x} = x/R_n$ | Axial coordinate from the cavitator base | m, - |
| $\xi = x/L_c$ | Normalized axial station along the cavity | - |
| $\boldsymbol{y}$ | Vector of true response | - |
| $\bar{y}$ | Mean of responses | - |
| $Z_d(.)$ | Gaussian random process of discrepancy | - |
| CFD | Computational fluid dynamics | - |
| EHK | Extended Hierarchical Kriging | - |
| FFD | Full factorial design | - |
| HF | High fidelity | - |
| LF | Low fidelity | - |
| Logvinovich I | empirical Logvinovich formula | - |
| Logvinovich II | analytical Logvinovich formula | - |
| PG-EHK | Physics-guided Extended Hierarchical Kriging | - |
| ẸHK | Extended Hierarchical Kriging | |
| HK | original Hierarchical Kriging | |
| RRMSE | Relative root mean square error | - |
| RMAE | Relative maximum absolute error | - |
| STD | Standard deviation | - |